\documentclass[12pt]{article}
\usepackage{amsmath,amssymb,amsfonts}

\usepackage[latin1]{inputenc}

\title{
The time at the subplanckian scale}
\author{\begin{tabular}{c}
{\sc C. Pierre\/}\\
\noalign{\vskip 6pt}
Institut de Mathématique pure et appliquée\\[-6pt]
Université de Louvain\\[-6pt]
Chemin du Cyclotron, 2\\[-6pt]
B-1348 Louvain-la-Neuve,  Belgium\\
pierre@math.ucl.ac.be\end{tabular}}
\date{\thispagestyle{empty}}

\def\To{\longrightarrow }

\begin{document} 

\begin{titlepage}

\thispagestyle{empty}
\maketitle

\begin{abstract}

With the theory of special relativity \cite{1}, time has been linked with space into a four-dimensional space-time from which a basic question must be asked: can space be really transformed into time and vice versa?  The response is affirmative if time has the same structural topological structure as space \cite{2} at the subplanckian quantum level in such a way that a discrete structural quantum time    constitutes the time part \cite{3} of the space-time internal vacuum of every elementary particle.  

It has thus been shown that a quantum time, quantized algebraically according to a lattice of time quanta, really exists and is emergent in the sense that  time quanta can be transformed into space quanta and vice versa.  Furthermore, this quantum time, only relevant at the subplanckian scale, is proved to be in one-to-one correspondence with the absolute and relative clock times.
\end{abstract}
\end{titlepage}

\section{Universal and relative clock times}

Understanding the nature of time has long been a main challenge for scientists \cite{4}--\cite{8} and philosophers \cite{9}.  Indeed, if, at first sight, it is relatively easy to realize that time has a continuous structure composed of a succession of instants corresponding to points on the line, a deeper insight of it reveals its close relationship with movement.  This is perhaps why Newton \cite{10} in his attempt to isolate time from the physical reality viewed it as a linear universal and absolute mathematical object.

Since  antiquity until Einstein, time was considered as being exterior to the physical phenomena \cite{11} allowing the measurement of their evolution.  This absolute time, conceived as a universal clock time,  measured the running flow of time and has been treated as a unidimensional object within the framework of the Euclidean geometry.

The physical reality was not only described by Newton with respect to a universal clock time but also in function of an absolute three-dimensional space: this gave him the absolute mathematical frame allowing to work out the dynamics of moving macroscopic  bodies.

In this classical dynamics \cite{12}, macroscopic objects move in the reference frame of the external three-dimensional space with respect to the absolute clock time.  Until special relativity, space and time were considered as independent concepts although mingled with the dynamics they describe.

It was only after the contribution of Minkowski, Lorentz, Poincare \cite{1} and principally Einstein \cite{13} to special relativity  that a four-dimensional continuous space-time world was introduced in Physics: it is sometimes called chronogeometry with reference to the clock time and is characterized by a Minkowski pseudo-Euclidean geometry \cite{1} in which the dimension of time is perpendicular to the three dimensions of space.  However, space and time are not completely equivalent in this context since time is not a fourth dimension of space.
\vskip 11pt

A basic concept of special relativity, allowing the description of the events by space-time coordinates $(x,y,z,t)$, is the inertial reference frame represented by a four-dimensional coordinate system moving at constant velocity.  The distance between two space-time events is obtained in the Minkowski-Lorentz geometry from their space-time interval which is the same in all inertial frames: this is equivalent to saying that the space-time interval between two events is an invariant of the transformation from an inertial frame into another given by the Lorentz transformations.

These considerations imply the two following drastic changes in our understanding of time:
\begin{enumerate}
\item the timelike (i.e. real) interval \cite{14} between two space-time events introduces the concept of proper time which is zero for the light in the vacuum in the frame of special relativity but different from zero for all other kind of matter, for example the elementary particles having a velocity inferior to the one of light $c$.

\item the absolute time (and space) of Newton \cite{10}, is replaced by a relative time (and space) associated with each reference body and varying with respect to its velocity and the other considered reference bodies.  In this respect, let us quote A. Einstein ``Every reference body has its own particular time; unless we are told the reference body to which the statement of time refers, there is no meaning in a statement of time of an event''.
\end{enumerate}
\vskip 11pt

Let then $s^2_{12}$ be this time like, i.e. causal, space-time interval between two events ``$1$'' and ``$2$'' \cite{14}
\[ s^2_{12}=c^2t^2_{12}-r^2_{12}>0\]
where the time and space coordinates of the two events are denoted respectively by $r$ and $t$ with $r^2=x^2+y^2+z^2$ in Cartesian coordinates $(x,y,z)$.

This concept of interval between two space-time events is the cornerstone of the Einstein-Minkowski-Lorentz-Poincare chronogeometry \cite{1} which is rigid, i.e. flat, in the frame of special relativity when it is curved in general relativity \cite{15}--\cite{17} due to the presence of matter deforming the neighbouring space-time. The clock time is thus affected by gravitational forces as it can be measured that a clock on the earth will run more slowly than a clock not subjected to gravitation.

Special and general relativity then tell us that every moving reference body, submitted to its related dynamics, has its own proper time:

\begin{enumerate}
\item[a)] calculated with respect to every external (moving) reference body to which the universal cosmic time can correspond;
\item[b)] deformed by (external) gravitational forces.
\end{enumerate}

A fundamental question is then the following:

Are space and time dependent in the frame of special (and general) relativity?
\vskip 11pt

The Lorentz transformation $\Delta t'=\Delta t\left(1-\frac{v^2}{c^2}\right)^{\frac12}$ shows that the (proper) time $\Delta t'$ of a body moving at velocity ``$v$'' is shorter than the corresponding time $\Delta t$ in the fixed inertial frame: in other words, a moving clock runs more slowly than a clock at rest.

Similarly, the Lorentz transformation $\Delta x'=\frac{\Delta x}{\left(1-\frac{v^2}{c^2}\right)^{\frac12}}$ gives the (proper) length $\Delta x'$ of our body moving at velocity $v$ with respect to the corresponding length $\Delta x$ in a fixed inertial frame: this shows that there is a dilation of lengths in the moving reference frame with respect to its equivalent at rest.

So, according to special relativity, a moving body experiences, in its own reference frame, time contraction and space dilation, with respect to the factor $\left(1-\frac{v^2}{c^2}\right)^{\frac12}$.

But, special relativity does not tell us that the contracted time has been transformed into dilated length: and, this seems a priori not possible since the clock time does not have the same topological structure as the three-dimensional space.

\section{Towards a quantum time of structure}

To obtain a new breakthrough in this problem, quantum (field) theory \cite{18} must be taken into account, and, especially its connection \cite{19} with special relativity.

A first remark can be  made: the time used in quantum theories refers mostly to the internal dynamics of the considered microscopic object and appears by means of the frequency ``$\nu \simeq \frac1{t_p}$'' inversely proportional to the period $t_p$ \cite{20}.  Now, this periodical time can be related to the astronomical time: there is thus a one-to-one correspondence between these two types of time.  For example, the frequency of the cesium atom is given in terms of a number of oscillations corresponding to one period of one second of astronomical time.

A new insight into this question will be reached if the reference frames of the special (and general) relativity are considered at the level of elementary particles.  Indeed, the energy of a particle is given in special relativity by:

\[ E=\frac{m\ c^2}{\sqrt{1-\frac{v^2}{c^2}}}\]
where $m$ is the rest mass of the particle.  For low velocity ($v\ll c$), the development of $E$ in power of $\frac vc $ is:
\[ E\approx m\ c^2+\frac{m\ v^2}2\]
where $E=m\ c^2$ is the rest energy of the particle.

This allows to find the famous relation:
\[ E^2=p^2\ c^2+m^2\ c^4\]
between the square of the energy $E$ of the particle and the square of its linear momentum $p$, which implies that:
\begin{enumerate}
\item[a)] if the particle is at rest, i.e. with $p=0$, then $E=m\ c^2$ gives the amount of energy which should be obtained by its annihilation in photon(s) from its rest mass.
\item[b)] $E\simeq p\ c$ is the amount of energy resulting from the annihilation of its rest mass $m$ (case $m \mathrel{\tilde\to} 0$)
into photons.
\end{enumerate}

Furthermore, if it is taken into account that, in quantum (field) theories, the following differential operators:
\begin{align*}
p_i &\To \frac\hbar i\ \frac\partial{\partial i} \qquad (\ i=x,y,z\ )\\
E &\To i\hbar\ \frac\partial{\partial t}\end{align*}
(and, by analogy, $m\to i\hbar\ \frac\partial{\partial t_0}$ where $t_0$ is the proper time)
\\
correspond respectively to the $i$-th component of the linear momentum and to the energy of our particle, it does not seem unwise to consider that the mass ``$m$" and the energy ``$E$'' or the linear momentum ``$p$'' are localized in two orthogonal spaces respectively of ``time''  and ``space'' type allowing the internal transfer of mass into energy and vice versa.

This is only possible if these two orthogonal spaces have a topological (spatial) structure as it was noticed before.

And, thus, we are forced to admit that elementary particles must be provided with an internal topological structure as it was introduced by the author \cite{3}.  Mathematics show that this internal topological structure must be of space (-time) type.  And, if we wish to connect \cite{22} quantum field theories (QFT) to general relativity, this internal space (-time) structure must be expanding in order that the cosmological constant \cite{21} of general relativity could correspond to it: it then would constitute the fundamental structure of the QFT vacuum shared out amongst the considered elementary particles.  In this perspective, every elementary particle would then be characterized by an internal expanding space-time structure constituting its own vacuum from which its matter shell could be generated.

It was justified in \cite{3} and in \cite{22} that the internal structure of an elementary particle must be:
\begin{enumerate}
\item[a)] composed of three embedded shells $ST \subset MG\subset M$: space-time ($ST$), middle-ground ($MG$) and mass ($M$).

\item[b)] of bilinear type, i.e. composed of the product of a left semiparticle localized in the upper half space by a symmetric right (co-)semiparticle localized in the lower half space.

Remark that the bilinearity was considered (contrary to QFT which works in a mathematical linear frame) in order to  really take into account the special relativity invariants in QFT.

\item[c)] really quantized algebraically in such a way that quanta then are irreducible algebraic closed subsets characterized by a Galois extension degree $N$.
\end{enumerate}

In the framework of the global program of Langlands \cite{23}, the time component of the most internal space-time shell ``$ST$'' of the vacuum of an elementary particle (mostly, fermion) was shown to be a time string field \cite{3}  composed of a double tower of packets  of symmetric strings behaving like harmonic oscillators and characterized by increasing integers labelling the number of quanta in these.  The time string field of a massive fermion can then generate, from singularities on it, the two covering middle-ground ($MG$) and mass ($M$) string fields in such a way that the latter corresponds to the proper mass structure.  The space string field of the ``$ST$'' shell of a fermion is orthogonal to the corresponding time string field and has the same type of structure.

Time and space string fields are thus structured according to a lattice of quanta and are emergent each one with respect to the other in the sense that time quanta can be transformed into space quanta by passing throughout the origin and vice versa.  

{\bf The time string field, which is thus a reservoir of space, is a discrete quantum time of structure\/} having an (anti)inertial effect on the corresponding quantum space field by slowing down the space-time field.

Every string at $n$ quanta of this time string field is characterized by:
\begin{enumerate}
\item a quantum periodical proper time $t_{0q}(n)$ which can be interpreted as a distance filled by a set of cycles referring to one quantum in this string like the wavelength.

\item a quantum frequency $\nu _q(n)$ which gives the number of oscillations per quantum in this time string.
\end{enumerate}

In order to connect this quantum time of structure to the clock times, two basic questions will finally be considered.  The first question can be stated as follows: is there a one-to-one correspondence between the discrete quantum time of structure and the universal clock time?  The reply is affirmative because to every time quantum of a string at $n$ quanta corresponds:
\begin{enumerate}
\item[a)] a period $t_{0p}(n)$ which gives the clock time needed for a set of cycles related to this quantum.
\item[b)] a periodical frequency $\nu _p(n)=\frac1{t_{0p}(n)}$ defined from $t_{0p}(n)$.
\end{enumerate}

The second question consists in knowing if 
a quantum time of structure can be defined with respect to
all types of clock times considered in this paper.  It is easy to realize that this will be the case if the quantum proper time $t_{0q}(n)$ and the quantum frequency $\nu _q(n)$ are dependent. 

Now, it was proved by the author \cite{3} in an attempt to solve the conjecture of Riemann \cite{24} that the trivial zeros of the zeta function $\zeta (s)$ \cite{25}--\cite{26}, related to the structure of strings at $n$ quanta (the integer $n$ varying), are in one-to-one correspondence with the equivalent non-trivial zeros of which imaginary parts $\gamma _n$ are approximately equal to the energies, i.e. the frequencies of these strings.  

It then results that the only relevant time at the subplanckian quantum scale is the quantum structural time of the space-time shell ``$ST$'' of the internal vacua of elementary particles in such a way that the universal and relative clock times be reduced to this quantum time at this microscopic scale.


\vskip 11pt

\begin{thebibliography}{99}

\bibitem{1}{} Lorentz, H., Einstein, A., Minkowski, H.    {\em The principle of relativity\/}, Dover, N.Y. (1923).

\bibitem{2}{} Atiyah, M.  Reflections on geometry and physics, {\em Surveys in Diff. Geom.\/}, {\bf 2\/}, 1--6 (1995).

\bibitem{3}{} Pierre, C.  {\em Algebraic quantum theory\/}, ArXiv math-ph/0404024 (2004).


\bibitem{4}{} Penrose, R.  {\em Structure of space-time in Battelle rencontres\/}, Ed. de Witt, B., Wheeler, J.A. Benjamin (1968).

\bibitem{5}{} 't Hooft, G.  {\em How does God play dice?  (Pre)-determinism at the Planck scale\/}, ArXiv-hep-th/0104219 (2001).

\bibitem{6}{} Hawking, S.W., Pennrose, R. {\em The nature of space and time\/}, Princeton Univ. Press (1996).

\bibitem{7}{} Hawking, S.W. {\em A brief history of time: From Big-Bang to black holes\/}, Bantam books, N.Y. (1988).

\bibitem{8}{} Jasselette, P. Le temps des physiciens, {\em Rev. Quest. Sci.\/}, {\bf 154\/}, 461--477 (1983).

\bibitem{9}{} Wang, W. Time in philosophy and in physics: from Kant and Einstein to Gödel, {\em Synthese\/}, {\bf 102\/}, 215--234 (1995).


\bibitem{10}{} Newton, I.  {\em Philosophiae Naturalis Principia Mathematica\/}, Londini, Jussu Societatis Regiae ac Typis J. Streater (1687).

\bibitem{11}{} Damour, T.  La relativité générale in: Qu'est-ce que l'Univers, {\em Université de Tous les Savoirs\/}, Vol. 4, dir. Y. Michaud, Ed. Jacob, O. (2000).

\bibitem{12}{} Arnold, V., Avez, A.    {\em Problèmes ergodiques de la mécanique classique\/}, Gauthier-Villars (1967).

\bibitem{13}{} Einstein, A.  Zur Elektrodynamik Bewegter Körper, {\em Ann. Phys.\/}, {\bf 17\/}, 891--921 (1905).

\bibitem{14}{} Einstein, A.   {\em Quatre conférences sur la théorie de la relativité\/}, Gauthier-Villars (1971).

\bibitem{15}{} Einstein, A.   {\em La théorie de la relativité restreinte et générale -- Exposé élémentaire\/}, Gauthier-Villars (1954).

\bibitem{16}{} Hawking, S.W., Ellis, G.F.R.   {\em The large scale structure of space-time\/}, Cambridge Univ. Press (1973).

\bibitem{17}{} Misner, C., Thorne, K., Wheeler, J.A.   {\em Gravitation\/}, Freeman.

\bibitem{18}{} Weinberg, S.   {\em The quantum theory of fields, Vols. I \& II\/}, Cambridge Univ. Press (1994).

\bibitem{19}{} Dirac, P.A.M.  Relativity and quantum mechanics, fields and quanta, {\bf 3\/}, 139--164 (1972).

\bibitem{20}{} Feynman, R.P., Leighton, R., Sands, M.   {\em The Feynman lectures on physics, Vol. 1\/}, Addison Wesley (1964).

\bibitem{21}{} Weinberg, S.  Einstein's mistakes, {\em Physics Today\/}, 31--36 (2005).

\bibitem{22}{} Pierre, C.   {\em A new track for unifying general relativity with quantum field theories\/}, ArXiv org, gr-qc/0510091 (2005).

\bibitem{23}{} Pierre, C.   {\em $N$-dimensional global correspondences of Langlands\/}, ArXiv org, math.RT/0510348 (2005).

\bibitem{24}{} Titschmarsch, E.C.   {\em The theory of the Riemann zeta function\/}, Oxford Univ. Press (1967).

\bibitem{25}{} Berry, M., Keating, J.  The Riemann zeros and eigenvalue asymptotics,   {\em Siam Rev.\/}, {\bf 41\/}, 236--266 (1999).

\bibitem{26}{} Katz, N. Sarnark, P.  Zeros of zeta functions and symmetry,   {\em Bull. Amer. Math. Soc.\/}, {\bf 36\/}, 1--26 (1999).

\end{thebibliography}
\end{document}